\documentstyle[aps,multicol,psfig,epsf,epsfig]{revtex}
\begin{document}

\def\xv{ {\bf x}}
\draft
\tightenlines

\title{Field theory of self-organized fractal etching}
   \author{Andrea Gabrielli $^{1,2}$, Miguel A. Mu\~noz$^{3}$, and 
Bernard Sapoval $^{2,4}$.}

\address{
$^1$INFM - Unit\`a di Roma 1 and Dip. Fisica, Universit\`a ``La Sapienza",
P.le A. Moro, 2, 00185 - Roma, Italy\\
$^2$Laboratoire de la Physique e la Matiere Condens\'ee, Ecole
Polytechnique- CNRS, 91128 Palaiseau, France\\
$^3$
Instituto de F{\'\i}sica Te\'orica y Compt. Carlos I,
Universidad de Granada, Facultad de Ciencias, 18071-Granada, Spain. \\
$^4$
 Centre de Math\'ematiques et de leurs Applications, Ecole
Normale Sup\'erieure - CNRS, 94140 Cachan, France\\
}                         

\date{\today}

\maketitle
\begin{abstract}

  We propose a phenomenological field theoretical approach to the 
chemical etching of a disordered-solid. The theory is based on a 
recently proposed dynamical etching model.  Through the 
introduction of a set of Langevin equations for the model evolution, 
we are able to map the problem into a field theory related to 
isotropic percolation. To the best of the authors knowledge,
it constitutes the first application of field theory to a 
problem of chemical dynamics. By using this mapping, many of 
the etching process critical properties are seen to be describable
in terms of the percolation renormalization group fixed point.
The emerging field theory has the peculiarity of being 
``{\it self-organized}'', in the sense that without any parameter 
fine-tuning, the system develops fractal properties up to certain 
scale controlled solely by the volume, $V$, of the etching solution.
 In the limit $V \rightarrow \infty$ the upper cut-off goes to 
infinity and the system becomes scale invariant. We present also a
finite size scaling analysis and discuss the relation of this 
particular etching mechanism with Gradient Percolation.
 Finally, the possibility of considering this mechanism as a new 
generic path to self-organized criticality is analyzed,
with the characteristics of being closely related to a real 
physical system and therefore more directly
accessible to experiments.
\end{abstract}

\pacs{PACS numbers: 64.60Ak, 81.65Cf}

\begin{multicols}{2}
\narrowtext

\date{\today}  

\section{INTRODUCTION}
   
   Corrosion of solids is an everyday phenomenon of
evident practical importance \cite{EU}. The recent development of theoretical 
tools for the study of disordered systems and fractals in the context of 
statistical mechanics \cite{Mandel,Meakin,Barabasi,HHZ} has triggered
an outburst of activity in this subject.

 When an etching solution is put in contact with a 
disordered etchable solid, the solution corrodes the weak parts of the solid 
surface, 
while the hard, stronger, parts stay uncorroded. 
During this process new regions 
of the solid, both hard and weak, 
are discovered and come into contact with the etching solution. 
As corrosion proceeds the etching power of the solution may
diminish: indeed, if the etchant is consumed in the reaction, etching becomes 
more and more unlikely until, finally, the solution is so impoverished and 
the solid surface so hardened that the 
corrosion process is arrested. 
At that moment all solid points in contact with 
the solution are too hard to be etched by the weakened etching solution. 
One of the most interesting aspects of this type of phenomenon is that the final 
solid-liquid interface has, in general, a fractal geometry, at least up to a 
certain scale \cite{Meakin,Barabasi,HHZ,kin}. This is precisely the qualitative 
phenomenology observed in a nice experiment on pit corrosion of aluminum thin 
films \cite{Balazs}.   

  Recently, a simple dynamical model of etching, 
capturing the aforementioned phenomenology, has 
been proposed \cite{model,GBS}. 
This model has been studied using both 
computational and analytical tools in \cite{GBS}, and from these 
studies strong evidence has been provided that the fractal properties of the 
solid surface, once the dynamics has stopped, are related to isotropic 
percolation. In principle, this is not an obvious result; in fact, at first 
sight, one could think that the interface should be anisotropic as there is a 
preferential direction in which the solution advances by etching the solid. 

   The purpose of this paper is to provide further theoretical evidence that 
indeed the critical behavior of the model dynamics is related to isotropic 
percolation. We also extend the previous relation  to spatial dimensions larger 
than $d=2$. To this aim, we shall first review (section 2) two known percolation 
models that will be useful in the forthcoming discussion: 
(i) dynamical percolation, and 
(ii) gradient percolation (GP).

Afterwards (section 3), we will define the dynamical etching model  
\cite{model,GBS} in a circular (spherical) geometry, and will derive a 
phenomenological field theory for it (section 4). From the analysis of this 
field theory the parallelism with percolation will be set up in a rather clear 
way, and this will provide further theoretical evidence on the connection
between etching and percolation phenomena.

The approach presented in this paper will allow us to study the system {\it 
self-organization} from a field theoretical point of view, and to verify that, 
in certain limit, the system is self-driven to the neighborhood of a critical 
point without need of any parameter fine-tuning. This is a new path to 
self-organized criticality \cite{BJP} as will be discussed in the last section.

\section{Two percolation models}

 In this section we review two different well-known percolation models that
will be useful in the discussion of the etching processes under consideration.

\subsection{Dynamical Percolation}
 Dynamical percolation is a model proposed for the study of the propagation of 
epidemics in a population, and/or for the analysis of forest fires. It is 
defined as follows \cite{epid,Jan}. Let us consider a regular square lattice; at 
each site there is a variable that can be in one of three possible states
(we borrow the language from epidemiology \cite{language}): (i) infected sites, 
(ii) healthy sites susceptible to be infected, and (iii) immune sites (non 
susceptible to be re-infected). At time $t=0$ a localized seed of infected sites 
is located at the center of an otherwise empty (healthy) lattice. The dynamics 
proceeds in discrete time steps either by parallel or by sequential updating
as follows: at each time-step every infected site can infect a (healthy) 
randomly chosen neighbor with probability $p$ or, alternatively, heal and
become immune to reinfection with complementary probability $1-p$. 
Any system state with no infected site is an {\it absorbing configuration}, 
i.e., 
a configuration in which the system can get trapped and from which it cannot 
escape \cite{Marro,Granada}. 
It is clear that depending on the value of $p$ the 
epidemics generated by the initial infection seed will either spread 
in the lattice (for large values of $p$) or die out (for small values of $p$). 
In all cases, the epidemics will leave behind a cluster of (healed) immune 
sites, infinite or finite respectively for the two aforementioned cases. 
Separating the two previous phases, there is a critical value of $p$, $0<p_c<1$, 
at which the epidemics propagates marginally, leaving behind a fractal cluster 
of immunized sites. 
It can be shown using field theoretical tools (see below) 
that this is a percolation cluster \cite{epid,Jan}. 
In this way we have a 
dynamical model which at criticality reproduces the (static) properties of 
standard percolation. 
Needless to say, the dynamical properties of the dynamical 
percolation equation, do not correspond to any known property of static 
percolation.

The dynamical percolation model can be cast into the following Langevin 
equation  \cite{epid,Jan} (or equivalently into a field theory 
\cite{BJW,Amit}): 
\begin{eqnarray}
\partial_t \rho(\xv,t) & =&  \mu \rho(\xv,t)  - \alpha
 \rho(\xv,t)
 \int_0^t dt' \rho(\xv,t')  \nonumber \\
 & + & \nabla^2 \rho(\xv,t) + \sqrt{\rho(\xv,t)} \eta(\xv,t)
\label{dyp}
\end{eqnarray}
where $\mu$ (the ``mass'' in a field theoretical language'')
 and $\alpha > 0$ are constants, $\rho(\xv,t)$ an activity field describing at a 
coarse grained level the density of infected sites, and $\eta(\xv,t)$ a Gaussian
white noise. Note the multiplicative nature of the noise, because of which
the state $\rho(\xv,t)=0$ defines an absorbing state, 
i.e. $ \partial_t 
\rho(\xv,t)=0$. 
Note also the presence of a non-Markovian term, that constitutes 
the key difference between this equation and the Reggeon field theory, 
characteristic of many other systems with absorbing states. 
This non-Markovian term stems from the existence of immunized sites, 
of which the system keeps indelible memory \cite{epid,Jan}.

The field theoretical and renormalization group analysis of Eq.(\ref{dyp}) can 
be found in the literature \cite{Jan}. 
The critical dimension is $d_c=6$, and 
the exponents, calculated in an epsilon expansion, coincide with the well known 
values for percolation calculated using other techniques \cite{Lub}. 
Apart from the static exponents, also a dynamical exponent $z$ can 
be derived from  this analysis of dynamical percolation \cite{Jan}.

\subsection{Gradient Percolation}

\begin{figure}[tbp]
\centerline{\psfig{file=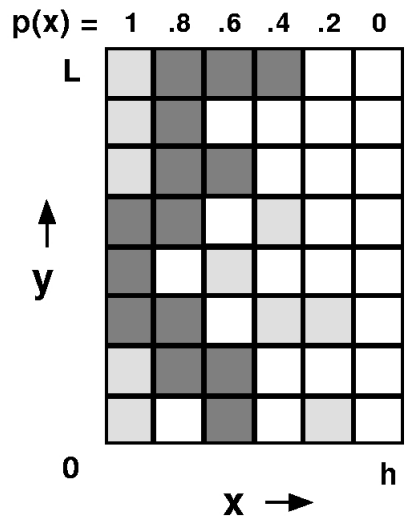,width=8cm,angle=-90}}
\caption{Schematic representation of the Gradient Percolation model.
In this case: $L=8$ and $h=5$. Grey (white) rectangles represent occupied 
(empty) sites. 
In darker grey we indicate the surface of the connected cluster of occupied 
sites. This surface has fractal dimension $D_f=7/4$ up to the 
characteristic thickness $\sigma\sim\nabla p ^{-1/D_f}$.}
\label{fig1}
\end{figure}

Gradient percolation \cite{GP} (GP) is defined in the following way: let us 
consider 
a bidimensional rectangular lattice of lateral sizes $L$ and $h$ respectively,
as shown in Fig.~\ref{fig1}.
 
An occupation probability given by $p(x)=1-x/h$ is assigned to  
sites in column $x$; this defines a transversal constant gradient,  $\nabla p = 
1/h$ for the occupation probability. 
Then, at each lattice site $(x,y)$ a random 
number $r(x,y) \in [0,1]$, extracted from an homogeneous distribution, and
representing the site reluctance to be occupied, is assigned. 
All sites with 
$r(x,y) < p(x)$ are declared occupied, while the remainings are empty. 
At the 
first column, $x=0$, all sites are occupied, while there is zero occupancy at 
the last one, $x=h$  (see Fig.~\ref{fig1}).
After identifying all sites as occupied or empty, one detects two connected 
regions (clusters): one (leftmost) with a majority of occupied sites possibly 
surrounding " lakes " of empty sites, and another one (rightmost sea) possibly 
surrounding islands. 
Separating these two regions there is an interface (the 
frontier of the connected cluster of occupied sites; it corresponds to the
 dark sites in figure 1). 
The average position of this interface can be 
shown to be at the square lattice site percolation threshold, $p_c$ 
\cite{GP,connect}. 
In fact, gradient percolation has been
 used as a computational tool to obtain accurate values of percolation
thresholds in different geometries by identifying the average position of the 
interface in sufficiently large lattices \cite{GP}. 
In the case that we are 
considering, the fractal dimension of the interface, $D_f =7/4$,
can be identified as the hull fractal dimension of the critical percolating 
cluster in a two-dimensional lattice \cite{connect}. 
There is an upper cut-off up to 
which this fractal behavior is observed; it is fixed by the width $\sigma$ 
which, in its turn, is determined by $h$, and therefore by $\nabla p$.
It can be shown using percolation theory that
\begin{equation} 
\sigma \sim {\nabla{p}}^{-\alpha_\sigma}
\label{ute}
\end{equation}
where $\alpha_\sigma = 1 / D_f$ \cite{GBS,GP,duplantier}.  
In order to have a well defined percolation system, with negligible 
finite size effects, the limit $L \gg \sigma$ has to be used. In this way, the 
length $h$, determining the value of $\sigma$, is the parameter that controls 
the finite size effects; the ``thermodynamic limit'' corresponds to $h 
\rightarrow \infty$ and $L\rightarrow\infty$ with both limits taken in the 
proper way \cite{GBS}. 
One can also estimate the variation of $p$ from 
on the leftmost to the rightmost extremes of the
wandering interface, $\Delta p $:
\begin{equation} 
\Delta p \sim {\nabla{p} }^{-\alpha_p}.
\label{ute2}
\end{equation}
The identity $\Delta p = \sigma*\nabla p = \sigma/h$
 imposes the following scaling relation among exponents:
 $\alpha_p = 1- \alpha_\sigma$, and therefore,
\begin{equation}
\alpha_p= { D_f -1 \over D_f}.
\end{equation}

Let us observe that gradient percolation can also be defined in a circular 
geometry, in which the gradient changes with the radial distance to 
the origin, and the cut-off is determined by the width of the roughly circular 
crown in which the interface is inscribed.

Summarizing, in this section, we have reviewed two well-known percolation 
models. 
Dynamical percolation is a model that, at its critical point, generates 
dynamically a percolation cluster. 
On the other hand, gradient percolation
is a static model, in which an interface appears with the same hull fractal 
dimension of the percolation cluster, but with no intrinsic dynamics defined.

\section{Dynamical etching model}

Having introduced the previous two percolation models, we go ahead by reviewing 
the dynamical etching model (DEM) at the focus of our study \cite{model,GBS}. 
It is defined by the following ingredients (see Figure 2).

(i) The random solid is mimicked by a two-dimensional square lattice of finite 
linear width $L$ and depth $Y$ ($Y$ can be arbitrarily large, or even infinite). 
Periodic boundary conditions in the finite direction are imposed, leading to a 
cylindrical geometry.

(ii) A random quenched number $r_i \in [0,1]$ (extracted from a uniform
distribution), assigned to each solid site $i$, represents 
the site resistance to etching.

\begin{figure}[tbp]
\centerline{
\psfig{file=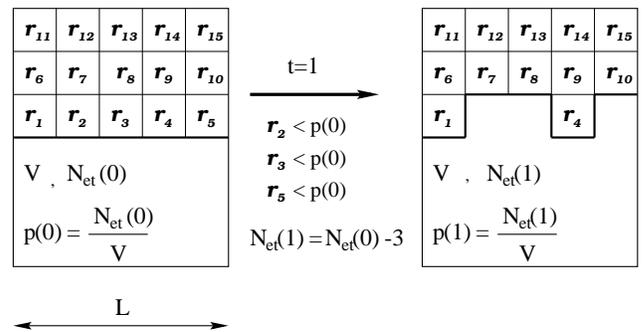,width=9cm}
}
\caption{Schematic representation of the dynamical etching model
in ``cylindrical'' geometry before and after the first time-step.
At this first time-step the (active)sites in contact with the solution are 
$i=1,2,3,4,5$, but only 2,3,5 have a resistance lower than the etching power
$p(0)$ and then are corroded. At the next time step new sites come in contact
with the solution (the whole second raw if the solution etches also in diagonal
direction). The etching power diminishes because of the consumption 
of etchant particles. Consequently sites $1,4$ stay uncorroded forever.}
\label{fig2}
\end{figure}

(iii) The etching solution occupies a fixed volume $V$ and is initially in 
contact with the solid through the bottom boundary, as depicted in 
Fig.~\ref{fig2}, defining a solid-solution interface advancing on average
in the upward direction.

The solution contains an initial number $N_{et}(0)$ of dissolved etchant 
molecules. 
Its concentration at time $t$ is  $C(t)=N_{et}(t)/V$. 
It is assumed 
that the etching power of the solution $p(t)$ is proportional to $C(t)$. 
Without
loss of generality the proportionality constant can be fixed to unity.
Following \cite{model}, we assume that etchant particles diffuse infinitely fast 
in the solution (at least  much faster than the characteristic time scale
of etching) and, hence, $p(t)$ is taken as spatially homogeneous, i.e., the  
etching power does not depend on the spatial position in the solution.

At each discrete time-step all solid sites located at the solid surface and 
satisfying $r_i < p(t)$ are dissolved (see Fig.~\ref{fig2}), i.e., they are 
removed from the solid, and a particle of etchant is consumed for each dissolved 
site, reducing in this manner the total etching power.
Calling $n(t)$ the number of dissolved solid sites (or equivalently the number
of consumed etchant particles) at time-step $t$, and $N(t)=\sum_{t'=0}^t n(t')$
the total number of etched solid sites up to time $t$, one can write
\begin{equation}
p(t+1)=p(t)-{n(t) \over V}= p(0)- {N(t) \over V}.
\label{pp}
\end{equation}

As $p(t+1) \le p(t)$, a site having endured the etching attack at time $t$ will 
also resist at any time $t' > t$ \cite{p0}. 
Furthermore, as a consequence of the 
corrosion process at time $t$, $m(t)$ new solid sites, previously in the solid 
bulk, come into contact with the solution at time $t+1$. 
Note that they are the 
sole candidates for corrosion at the next time-step. 
Finally, since the solution 
has the possibility to detach finite solid islands, the global solid surface 
is composed 
both by the surfaces of the detached islands, and by the set of solid sites
separating 
the solution from the bulk. 
This interface is called the {\em corrosion front}. 
A more detailed description of the model phenomenology can be 
found in \cite{GBS}. 
Here we simply summarize
the main features of the corrosion front at the arrest time $t_f$.
They are well represented by GP with $\nabla p\sim L/V$: 

(i) the corrosion front shows fractal features with
$D_f\simeq 1.75$ up to a characteristic scale (front thickness) $\sigma$;

(ii) $\sigma\sim (L/V)^{-1/D_f}$;

(iii) $p_c-p(t_f)\sim (L/V)^{-\alpha_p}$, with $\alpha_p\simeq (D_f-1)/D_f$
(therefore, in the right thermodynamic limit $p(t_f)\rightarrow p_c$).

\begin{figure}[tbp]
\centerline{
\psfig{file=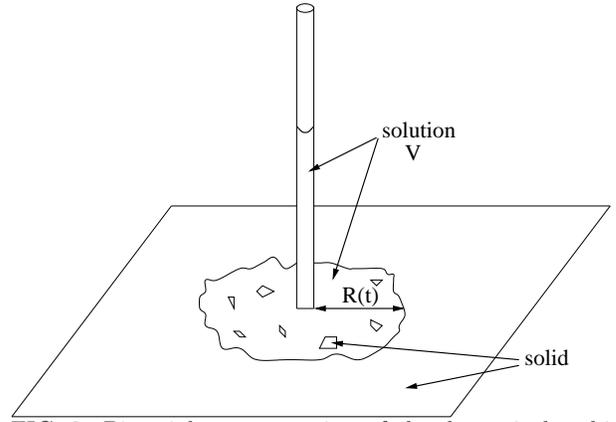,width=8cm}
}
\caption{Pictorial representation of the dynamical etching model in
``spherical'' geometry.}
\label{fig3}
\end{figure}

Let us introduce here a slight geometrical modification of the DEM which makes 
more clear the connection to dynamical percolation. Instead of considering
a cylindrical geometry with the etchant solution invading the cylinder
from  the bottom (as in Fig.~2), we consider a flat infinite lattice,
in which the volume $V$ of the etching solution is poured at time $t=0$ at an 
arbitrarily chosen central site as schematically shown in Fig.~\ref{fig3}. The 
volume $V$ of the etching solution is constant. 
Observe that with this geometry 
the model has some clear analogies with dynamical percolation. 
The main difference is that, in the spherical DEM, the control 
parameter (the corroding or infecting probability) is not a constant but 
decreases in time as the etching process proceeds. As in cylindrical 
geometry, the dynamics can be roughly divided into two regimes \cite{GBS}:
a {\em smooth} one when $p(t)$ is much larger than $p_c$, and a {\em critical} 
one when $p(t)$ approaches $p_c$.
In the smooth regime, fluctuations around the average behavior are small 
while in the critical regime 
fluctuations dominate the dynamics \cite{model}. 
Indeed, at early time-steps, the etching power being  sufficiently
larger than $p_c$, 
it is simple to show \cite{GBS} that the corrosion front
is an approximate expanding
circumference centered at the origin, and the  number
of new solid sites coming into contact with the solution
at time $t$ satisfies the approximate relation
$m(t)\simeq 2 ~ \pi ~ R(t)$,
where $R(t)$ is the maximal radius reached by the corrosion
up to time time $t$.
As the etchant power is reduced, the
corrosion front becomes rougher and rougher, until the dynamics
is finally arrested (see Fig.~\ref{fig3}).
Since in the smooth regime $m(t)\gg 1$,
we can write $n(t) \simeq p(t)m(t)$.
Hence, within this approximation,
it is possible to write down the following equation:
\begin{equation}
p(t+1)\simeq p(t)- \frac{2 \pi R(t) p(t)}{V}.
\label{smooth}
\end{equation} 
Since in this
regime obviously $R(t) \propto t$, we can write Eq.~\ref{smooth}
in a differential form as follows:
\begin{equation}
{d p(t) \over d t} \simeq - {2 ~ \pi ~ t \over V} p(t)\,,
\end{equation}
whose solution is
\begin{equation}
p(t) \simeq  p(0) \exp{ \left( -\pi t^2 /V \right)}.
\label{exp}
\end{equation}

   From this, the characteristic time of the 
dynamics is seen to be proportional to 
$\sqrt{V}$; i.e. the etching power of the solution reaches the value $p_c$ 
in a time $t_c$ proportional to $\sqrt{V}$. 
Moreover, as $R(t) \propto t$, at the 
cross-over between the smooth and the critical regime (i.e. when $p(t)\simeq 
p_c$) the solution reaches a distance $R_c\simeq \sqrt{V}$ from the origin.  
By differentiating this expression we conclude that the gradient of values of 
$p$ at which different sites have been corroded in the radial direction is 
proportional to $R/V$. 
Finally, as $R(t_c) \propto \sqrt{V}$, the gradient $\nabla p$ at 
$t_c$ is proportional to $1/ \sqrt{V}$. 
In this way, in analogy with the cylindrical case, 
we expect that the geometrical properties of the final corrosion front 
are well represented by GP where the gradient of $p$ is 
dynamically generated. 
Replacing $\nabla p$ with $1/ \sqrt{V}$, the scaling 
relations studied for gradient percolation can be extended 
to the present case.
The previous description is valid only for the smooth regime, i.e.
up to the time at which $p(t) \approx p_c$.
However, since the critical regime
is shorter than the smooth one, we have $t_f \sim 
R(t_f)\sim \sqrt{V}$, where $R(t_f)$ is the average radius of the final 
corrosion front, and the previous estimations remain valid. 
In order to check that also in the critical 
regime the radial gradient of the solution etching power is 
given by $R/V$, 
it is sufficient to assume that during this regime the corrosion 
front changes from a quite smooth geometry to a rougher one, with a final 
thickness $\sigma$.
Because of the much shorter duration of the critical regime 
one has $\sigma\ll R(t_f)$. 
In this way, during the critical regime the solution 
etches a number of solid sites proportional to $\sigma R(t_f)$. 
Therefore, from 
Eq.~(\ref{pp}), the variation of the etching power in this regime 
in average is $\Delta p\sim \sigma R(t_f)/V$.

In conclusion, we have defined a spherical version of the DEM, and seen its 
connection with gradient percolation: given the time-diminution of $p$, the 
system generates dynamically a spatial gradient of the values of 
$p$ at which the different sites were etched. 
Let us finally emphasize that if, after the process is arrested, 
more etchant solution is added then the process continues until it
is stopped again at a value of  $p$ around $p_c$. In this way the
disordered solid plays the role of a chemical buffer.
In the next section we present a more 
theoretical treatment allowing us to draw even more precise connections 
between DEM and percolation theory.

\section{Phenomenological field theory}

In order to construct a field theoretical description for the
dynamical etching model, a possibility  would be to write down the master 
equation defining the dynamics and then (using a Poissonian transformation 
\cite{Gardiner,Yo} or alternatively a Fock-space formalism \cite{Peliti}), 
derive a generating functional\cite{Amit}.
Instead of following that strategy, we 
prefer here to present a phenomenological set of stochastic 
Langevin equations describing the model at a mesoscopic scale. 
This direct approach, ``\`a la 
Landau'', based on the analysis of the main symmetries and conservation 
laws of the discrete model, 
has proven very efficient in describing many other systems related to 
percolation, directed percolation and, in general, 
systems with absorbing states \cite{Jan,Granada}.  

Let us consider the following three different local densities
or coarse-grained fields:
\begin{itemize}
\item
$s(\xv,t)$ describing the local density of 
material susceptible to be etched at any time after $t$. 
In the discrete model there are two types of sites contributing
to this density: 
(i) bulk solid sites and (ii) ``fresh'' interface sites i.e.,
solid sites freshly arrived to the solid-liquid interface 
(susceptible to be etched at the next time-step).
\item $q(\xv,t)$ is the local density of passivated and inert material.
In the microscopic model this is the density of interface 
sites having already resisted an etching trial, i.e., 
immune or not-susceptible to be corroded at any future 
time-step. 
\item $c(\xv,t)$ is the local density of corroded and replaced by solution 
sites, i.e., the local
 density of etchant.

\end{itemize} 
The mean field equations (rate equations) describing the evolution of the 
averaged mean values of these magnitudes are:
\begin{eqnarray}
\dot{s}(t) & = & -  \alpha         c(t) s(t) \nonumber  \\
\dot{q}(t) & = &    \alpha (1-p(t)) c(t) s(t)  \nonumber \\
\dot{c}(t) & = &    \alpha  p(t)    c(t) s(t)  
\label{deterministic} 
\end{eqnarray}
where $p(t)$ is the probability to etch an active site at time $t$, 
and $\alpha$ is a positive constant. 
In what follows, and without loss of generality we fix $\alpha=1$.
The interpretation of the first equation is: in order for the density of 
susceptible sites to change (decrease) in a region, it is necessary to have 
locally both a non-vanishing density of etchants and raw solid material 
susceptible to be etched. 
This restricts the dynamics to {\it active} regions, 
i.e., zones in the interface separating the 
etchable-solid and the solution, in 
which non-vanishing local densities of $s$ and of $c$ coexist. 
Moreover, the second and the third relations in Eq.~\ref{deterministic} 
express the fact that an active site becomes either a $c$-site, 
with probability $p(t) $ (the corrosion 
power at time $t$), or alternatively, after healing, a $q$-site with 
complementary probability $1-p(t)$. 
Note that, as $\dot{s}~+~\dot{c}~+~\dot{q}~=~0$, 
the total number of sites is conserved during 
the dynamics. 
It is worth stressing that Eq.~(\ref{deterministic}) captures the 
fact that sites resisting an etching attempt remain un-corroded indefinitely
(as occurs in the microscopic model). 
In fact, the number of $q$-sites grows 
monotonously until the etching process is arrested. 
 
Observe that we have written so far mean field equations in which spatial 
dependence and fluctuations are not taken under consideration. 
At this point, it is convenient to introduce an activity field 
$\rho(\xv,t)\equiv c(\xv,t)s(\xv,t)$ (or $\rho(t)\equiv c(t)s(t)$ 
as long as  spatial dependences are omitted).  

   From Eq.~\ref{deterministic} it follows immediately that
\begin{equation} 
\dot{\rho}(t) = - c(t) \rho(t) + p(t) s(t) \rho(t) \,.
\label{det2}
\end{equation}

In order to implement in our theoretical description 
the diminution of the etching power as the corrosion process proceeds, 
we write $p(t)$ (analogously to 
Eq.(4)) as
\begin{equation}
p(t)= p_0 - {N(t) \over V}\,,
\label{p}
\end{equation}
where now $N(t)=\int d \xv [c(\xv,t)-c(\xv,0)]$ is the number of consumed 
etchant particles up to time $t$ and, as before, $V$ is the solution volume.
Integrating in time the equation for $c$ in Eq.~\ref{deterministic}, 
integrating 
then in space, and substituting the result into Eq.~\ref{p}, we obtain
\begin{equation}
p(t)=p(0) \exp\left[ - {1 \over V} \int_0^t dt'
 \int d\xv' \rho(\xv',t') \right].
\label{p3}
\end{equation}
 Plugging this result into Eq.~\ref{det2}, and re-introducing the 
spatial dependence of the fields, it is a matter of simple algebra to obtain
\begin{eqnarray}
 \partial_t \rho(\xv,t) & = & [ p(t) s(\xv,0) - c(\xv,0)]
 \rho(\xv,t)   \nonumber \\ 
 & - & \rho(\xv,t) \left[ \int_0^t dt' (1-p(t')) \rho(\xv,t') \right].
\label{det3}
\end{eqnarray}
In order to go beyond this mean field description, it is necessary 
to include properly spatial coupling and fluctuations (as a noise).
\begin{itemize}
\item {\it Spacial coupling} \\
 In principle, the spatial coupling can be taken into account by
introducing additional terms into Eq.~\ref{det3}. 
However, because of the 
isotropic nature of the local dynamics, terms not invariant under space 
inversion (as odd derivatives of the fields) are not allowed. 
Also, terms like 
$|\nabla\rho(\xv,t)|^n$ cannot appear, given the absence of surface tension in 
the microscopic rules \cite{Barabasi}. 
Therefore, only terms as $\nabla^{2n} \rho(\xv,t)$ 
and higher powers of them are allowed. We introduce in 
Eq.~\ref{det3} the only relevant term in the renormalization group optics 
\cite{Amit}, namely a diffusive coupling $\nabla^2 \rho(\xv,t)$, that 
typically appears in continuous descriptions of interacting particle systems. 
It will be checked {\em a posteriori} that, as a matter of fact, the omitted 
terms are irrelevant at criticality.
\item {\it Noise} \\
 In order to introduce properly the noise term, 
let us consider a small region in 
which there are $k$ ``fresh'' surface sites in contact with the solution of 
etching power $p$. 
Since these fresh sites have random independent resistances,
the average number of dissolved sites will be $p \cdot k$ 
and the fluctuation from this average
number will be Poissonian, i.e. of order $\sqrt{p\cdot k}$.
This implies that in the continuous description fluctuations 
of $\rho(\xv, t)$ are proportional to its square-root. 
Consequently, a term $\sqrt{\rho(\xv,t)} \eta(\xv,t)$ 
has to be added to Eq.~\ref{det3}, with $\eta(\xv,t)$ being
a Gaussian white noise with zero mean and no spatio-temporal 
correlations: $ < \eta(\xv,t) \eta(\xv', t') > = 
\delta(\xv-\xv') \delta(t-t')$. 
Deviations from gaussianity 
and higher order corrections can be easily argued to be irrelevant in the 
renormalization group optics, and therefore are not taken into account. 
This type of noise, with amplitude proportional to 
the square-root of the activity field, 
is characteristic of systems exhibiting a transition from an active to an 
absorbing phase \cite{Yo}: 
let us emphasize that wherever the activity field is 
zero, the dynamics is stopped \cite{Marro,Granada}.
\end{itemize}
Introducing these two new ingredients in Eq.\ref{det3} one obtains finally
\begin{eqnarray}
 \partial_t \rho(\xv,t) & = & [ p(t) s(\xv,0) - c(\xv,0)]
 \rho(\xv,t)   \nonumber \\ 
 & - & \rho(\xv,t) \left[ \int_0^t dt' (1-p(t')) \rho(\xv,t')\right]   
\nonumber \\  &+& \nabla^2 \rho(\xv,t) + \sqrt{\rho(\xv,t)} 
 \eta(\xv,t)
\label{ext}
\end{eqnarray}
up to higher order, irrelevant, terms.

It is worth stressing that even though the microscopic model has originally 
quenched disorder, 
it has been possible to describe it in terms of a stochastic 
equation with annealed noise. 
This simplification owes to the fact that every 
site is tested for corrosion at most once. 
If it survives, it will stay un-corroded indefinitely, 
as previously explained. 
In this way, as every random number is used at most once it has 
not to be stored, there are no time  
correlations in the noise and, consequently, a stochastic process with no 
quenched disorder can be used to cast the discrete model.

It is also important to underline that wherever $\rho(\xv,t)$ vanishes, all 
activity, including fluctuations, ceases in Eq.\ref{ext}. 
In other words, 
$\rho(\xv,t)=0$ defines an absorbing state \cite{Granada,Marro}. 
This is just 
another way of saying that in the microscopic model, whenever there is no 
contact between the etchable solid and the solution, i.e., when they are 
separated by an interface of passivated (immune) solid sites, 
the dynamics is arrested. 
Continuous descriptions of systems with absorbing states are based 
on equations for the activity. 
In the neighborhood of any absorbing state phase
transition the activity is small, 
and series expansions on the activity density 
(as the ones we have used to arrive at Eq.\ref{ext}) are justified 
\cite{Marro,Granada}.

Observe that, apart from the time dependence, of $p(t)$, Eq.~\ref{ext} is
identical to the Langevin equation describing {\it dynamical percolation} 
(see Eq.~\ref{dyp}) \cite{epid,Jan}. 
As described above, dynamical percolation 
is a rather well known dynamical process generating 
percolation clusters with a characteristic size determined by $\mu$ 
(see Eq.~\ref{dyp}). 
Let us analyze the differences between Eq.~\ref{ext} and Eq.~\ref{dyp}. 
Particular attention 
must be paid to the exponential factor in the expression for $p(t)$ (see 
Eq.~\ref{p3}) absent in dynamical percolation; due to it some of the
coefficients in Eq.~\ref{ext} are time dependent, while their counterparts in 
Eq.~\ref{dyp} are constants. 
First we discuss whether this extra time-dependence
 may affect the critical properties. 
Note that $p(t)$ does not fluctuate (as verified in simulations in \cite{GBS})
because it is a smooth function of the integral of the field over all the past 
time and the whole space. 
Therefore, it is a deterministically decreasing
time-dependent term; 
or in other words it depends on spatio-temporal integrals 
of the activity field and not on the local activity field itself. 
Hence, this term has not critical fluctuations, 
and does not affect the system critical properties. 
However, it is crucial in order to characterize the temporal 
crossover from the active to the absorbing phase. 
Indeed, as the linear term 
coefficient in Eq.~\ref{ext} includes a dependence on $p(t)$, 
for early times it is positive, corresponding to the fact that 
for early times the system is in the supercritical regime. 
As the etching mechanism proceeds, the argument of the 
exponential in Eq.~\ref{p3} grows in modulus, and there is a finite time, 
$t_c$, at which the linear term coefficient takes its critical value. 
Immediately after, the process becomes subcritical \cite{mass}, 
i.e. it reaches the absorbing phase and, the dynamics tends 
to be stopped with an exponentially fast rate (i.e. the number 
of active sites decreases exponentially).
Furthermore, as more and more sites are etched, the linear 
coefficient in Eq.~\ref{ext} becomes smaller and smaller than its critical 
value, and the exponential stopping rate is accelerated 
till a final time $t_f$, 
at which an absorbing (blocking) configuration is reached. 
Therefore, the main effect of the time-dependent linear term 
coefficient is that, by continuously 
diminishing, it drives the system to the neighborhood of the
dynamical-percolation-field-theory critical point, 
but it does not add any new relevant operator that 
could eventually change the universality class.  
At this point, it is a matter of simple algebra to verify
that all the terms omitted in our derivation are indeed irrelevant at the 
dynamical percolation, Eq.~\ref{dyp}, renormalization group fixed point.

Given the previous discussion, one is allowed to say that Eqs.~\ref{p3} 
and \ref{ext} define a {\it self-organized dynamical percolation Langevin 
equation}: without tuning any parameter the dynamics is arrested in the 
neighborhood of the percolation critical point, and critical (fractal) 
properties can be measured up to certain length scale 
determined solely by the 
parameter $V$ (which controls the decrease rate of $p(t)$ ).
In the limit $V \rightarrow \infty$ the upper cut-off goes
to infinity; i.e.,  for sufficiently 
large values of $V$ the point at which the dynamics is 
stopped occurs at values 
of $p$ arbitrarily closed to its critical value in Eq.~\ref{dyp}.

Observe that the microscopic process is also self-organized: 
initially $p(t)$ is 
taken to be super-critical, but it decreases monotonously till it reaches a 
critical value, and as soon as the sub-critical regime is reached 
the process is stopped in an exponential way. 
Therefore, our continuous description reproduces 
the essential features of the microscopic model.

The long range correlations (generating fractal behavior) in the dynamical 
etching process are generated in the regime in which the linear coefficient 
takes values around its critical value. 
As a consequence, it is inferred that
the fractal properties of the final frozen configuration are related to the 
standard dynamical percolation renormalization group fixed point
in any dimension. 

Up to higher order terms, $q(\xv,t)$  can be written as 
$q(\xv,t) \propto \int dt' \rho(\xv,t)$. 
This variable, the integral over the past history of the 
activity field, represents the  statistics of immunized sites as described by 
Janssen \cite{Jan}, and therefore, in our problem the statistics of 
``surviving'' solid clusters in \cite{model,GBS} is also 
related to percolation properties. 
Observe that regions of the cluster of corroded sites 
far from the final blocking corrosion front have been corroded with a 
value of $p$ larger than its critical value and therefore are not critical.
For the same reason the final corrosion front can be seen as
the external perimeter of an invading percolation cluster (the
etchant solution) with $p\simeq p_c$. 
This explains the value of the fractal 
dimension $D_f\simeq 1.75$ found for this final corrosion front,
which is nothing but the hull exponent of 
percolation in the lattice geometry under 
consideration \cite{GBS,duplantier}. 

Finally, we can perform a finite size scaling analysis of our equation in order 
to determine how different magnitudes scale as a function of the only free 
parameter, $V$; in particular, we can evaluate the distance from the critical 
point at which the process will be finally stopped. 
The linear coefficient in Eq.~\ref{ext} is 
\begin{eqnarray}  
\mu(\xv,t) & = & s(\xv,0) p(0) \exp \left[-{1\over V} \int_0^t
d t' \int d \xv' \rho(\xv',t') \right] \nonumber \\ 
 & - & c(\xv,0). 
\label{masa}
\end{eqnarray}
For the points where the dynamics is arrested, at $t_f$, 
it is clear that $s(\xv,0)=1$ and 
$c(\xv,0)=0$; i.e. at $t=0$ they belong to the solid bulk. 
Hence, for the bulk 
we have $\mu(t) = p(0) \exp \left[ -{1\over V} \int_0^t d t' \int d \xv' 
\rho(\xv',t') \right]$ independent of $\xv$. 
As usual $t_c$ is the time at which the 
process is critical  $\mu(t_c)=\mu_c$, and $t_f$ the time at which the
process is actually arrested. 
For large values of $V$ the argument of the exponential in Eq.~\ref{masa}
can be expanded in power series, and we have, up to the
leading order: 
\begin{equation}
\mu(t_c)-\mu(t_f) \propto {1 \over V} 
\int_{t_c}^{t_f}
d t' \int d \xv' \rho(\xv',t').
\label{plof}
\end{equation}
Now we can  use the scaling analysis presented in the first 
part of the paper. 
In the time interval between $t_c$ and $t_f$ the solution erodes 
a region of the solid with, roughly speaking the shape of a circular hole
limited by a crown of radius $R_c$ and width $\sigma$. 
As argued before $R_c \propto \sqrt{V}$.
Moreover, since the quantity 
$\int_{t_c}^{t_f} d t' \int d \xv' \rho(\xv',t')$
gives the amount of material tested by the etching solution between 
$t_c$ and $t_f$, we can write
$\int_{t_c}^{t_f} d t' \int d \xv' \rho(\xv',t') \sim R_c ~ \sigma$.
Therefore $\mu(t_c)-\mu(t_f) \propto \sigma/\sqrt{V}$. 
Writing $\sigma \propto (1/\sqrt{V})^{-\alpha_\sigma}$ one obtains
\begin{equation}
\mu(t_c)-\mu(t_f) \propto \left(\sqrt{V}\right)^{\alpha_\sigma-1}\,.
\label{d-mu}
\end{equation}

The fact that $V$ is finite then implies that $\sigma$ is 
finite and $\mu(t_c)- \mu(t_f)\ne 0$. 
It is only in the large $V$ limit that the process is stopped 
exactly at the dynamical percolation critical point. 
As discussed in section 3, 
one has $\alpha_{\sigma}=1/D_f$ \cite{GBS,duplantier}, and therefore the 
distance from the final mass to the critical one scales as $\Delta p$, 
that is as the excursion of the occupancy probabilities 
along the separation interface of GP.
Therefore we have determined not only the 
universality class, but also established how finite size effect operate in 
the DEM.

In conclusion, all the critical (fractal) properties of the 
microscopic model can be shown to be related to (dynamical) 
percolation in any space dimension, by 
using the above presented continuous (field theoretical) 
representation, and finite size corrections can be evaluated.

\section{Conclusions}

Summing up, we have found that the dynamical etching model proposed by Sapoval 
et al., is a self-organized process describable by a the continuous Langevin 
equation similar to that of dynamical percolation. This Langevin equation 
includes a linear term coefficient that decreases monotonously as the etching 
process goes on. In this way as soon as it takes it critical value and
enters the ``absorbing'' regime the etching process is stopped. 
Consequently, the fractal (scale invariant) properties of the 
interface in the etching process are shown to be related 
to the (dynamical) percolation renormalization group fixed point. 
This result is valid in any space dimension. In particular, our 
analysis permits us to conclude that the etching  upper critical dimension is 
$d_c=6$, as in (dynamical) percolation. 
On the other hand, we have also evaluated the role of finite 
size corrections that are essentially different from 
those of standard dynamical percolation.

An interesting aspect from a theoretical perspective is that 
the field theory (Langevin equation) describing the process 
is self-organized, in the sense that, without any parameter 
fine-tuning, fractal, scale-invariant properties are generated. 
However, it is only in the limit $V \rightarrow \infty$ ($1/V 
\rightarrow 0$), that the upper cut-off for scaling diverges. 
This is in clear analogy with what occurs in other models of 
self-organization, as sandpiles \cite{BTW} for which critical 
behavior is observed in the limit of dissipation and driving 
going to zero \cite{FES}.

The mechanism discussed in this paper constitutes a new ``path'' 
to self-organized criticality \cite{BJP}, 
in which the control parameter decreases 
monotonously until it reaches the neighborhood of the absorbing-state
phase transition at which the dynamics is 
arrested in an exponential way. 
This same mechanism will be investigated in the context of different 
types of absorbing-state phase transitions 
(as directed percolation \cite{Marro,Granada})
in a future work. 
Observe that this scenario has the great advantage of being
related in a clear-cut way to real physical systems, making
therefore the observation of self-organized criticality much
more accessible to experiments.

\vspace{1cm}
{\centerline{\bf ACKNOWLEDGMENTS}}
We acknowledge useful discussion with A. Baldassarri, U. Marini Bettolo Marconi 
and P. Hurtado. We acknowledge partial support from the 
European Network contract 
ERBFMRXCT980183,and DGESIC (Spain) project PB97-0842. 
We thank It. FFSS for 
giving us the opportunity to discuss this problem.
M.A.M. acknowledges the kind hospitality at the Ecole Polytechnique where this 
work was started.

\end{multicols}
 

\begin{references}

\bibitem{EU} U. R. Evans, {\it The corrosion and Oxidation of Metals:
Scientific Principles and Practical Applications},
Arnold, (London), 1960). H. H. Uhlig, {\it Corrosion and 
Corrosion Control}, (Wiley, New York), 1963.

 \bibitem{Mandel} B. Mandelbrot, {\it The fractal Geometry of Nature},
(Freeman, San Francisco, 1982).

\bibitem{Meakin} P. Meakin, {\it
Fractals, scaling and growth far from equilibrium},
Cambridge Nonlinear Science Series 5,
(Cambridge University Press), 1998.
 
\bibitem{Barabasi} A. L. Barabasi and H. E. Stanley, {\it Fractal
Concepts in Surface Growth}, (Cambridge University press,
Cambridge, 1995). 

\bibitem{HHZ} T. Halpin-Healy and Y.-C. Zhang, 
Phys. Rep. {\bf 254}, 215 (1995).


\bibitem{kin} {\it Percolation Structures and Processes},
 edited by G. Deutscher, R. Zallen, and J. Adler,
Annals of the Israel Physical Society,
Vol. 5 (Adam Hilger, Bristol), 1983.
          
       
\bibitem{Balazs}
 L. Bal\'azs, Phys. Rev. E {\bf 54}, 1183 (1996).

\bibitem{model} B. Sapoval, S. B. Santra, and P. Barboux,
  Europhys. Lett., {\bf 41}, 297, (1998).
S. B. Santra and  B. Sapoval, Physica A {\bf 266}, 160 (1999).

\bibitem{GBS} A. Gabrielli, A. Baldassarri, and B. Sapoval,
Phys. Rev. E. {\bf 62}, 3103 (2000).

\bibitem{BJP} R. Dickman, M. A.  Mu{\~n}oz, A. Vespignani, 
and S. Zapperi, Brazilian J. of Physics {\bf 30}, 27 (2000).    

\bibitem{p0} Observe that at early times $p(t)$ has to be
larger that percolation threshold $p_c$; otherwise the
dynamics comes to its end without reaching a critical
regime.
    
\bibitem{epid}
J.L. Cardy, J. Phys. A {\bf  16}, L709 (1983);
J.L. Cardy and P. Grassberger, J. Phys. A {\bf 18}, L267 (1985).
 
\bibitem{Jan}
H.K. Janssen, Z. Phys. B {\bf 58}, 311 (1985).            

\bibitem{language} In the language of excitable media
(autocatalytic chemical reactions, propagation of
electrical activity in neurons, spreading of diseases)
there are, in general three types of states called:
quiescent (healthy), excited (infected), and refractory
(immune).
 

\bibitem{Marro}  J. Marro and R. Dickman,
 {\em Nonequilibrium Phase Transitions in Lattice Models},
(Cambridge University Press, Cambridge, 1999).        
 
\bibitem{Granada} See, G. Grinstein and
 M. A. Mu\~noz, {\it The Statistical
Mechanics of Systems with Absorbing States
}, in ''Fourth Granada Lectures in Computational Physics'',
edited by P. Garrido and J. Marro,
Lecture Notes in Physics, Vol. 493 (Springer, Berlin 1997),
p. 223, and references therein. See also, H. Hinrichsen, Adv. Phys. 
{\bf 49}, 815 (2000).            


\bibitem{BJW} R. Bausch, H. K. Janssen, and H. Wagner,
Z. Phys. B {\bf 24}, 113 (1976). C. De Dominicis and 
L. Peliti, Phys. Rev. B {\bf 18}, 353 (1978).


\bibitem{Amit} D. J. Amit, {\em Field Theory, the Renormalization Group
and Critical Phenomena}, (World Scientific, Singapore, 1992).
  J. Zinn-Justin, {\it Quantum Field Theory and Critical Phenomena},
(Clarendon Press, Oxford, 1990).              


\bibitem{Lub} T. C. Lubensky and J. Isaacson, Phys. Rev. Lett. {\bf 41},
829 (1978); Phys. Rev. A{\bf 20}, 2130 (1979).
 
\bibitem{GP}
B. Sapoval, M. Rosso, and J. F. Gouyet,
 J. Phys. Lett. (Paris), {\bf 46}, L149 (1985);
B. Sapoval, M. Rosso, and J. F. Gouyet, in {\em The Fractal Approach
to Heterogeneous Chemistry}, edited by D. Avnir (John Wiley and Sons Ltd.,
New York, 1989). M. Rosso, J. F. Gouyet, and B. Sapoval,
Phys. Rev. B {\bf 32}, 6053 (1985).

\bibitem{connect} R. Ziff and B. Sapoval, J. Phys. A : Math. Gen. 19, L1169-1172 
(1986). The connectivity criterion is a relevant ingredient to be taken in 
consideration. For example, in a square lattice, the connection between occupied 
sites is to first neighbor while the connection between empty (etched) sites is 
to first and second nearest neighbors. See the discussion in \cite{GBS}.

                                     
\bibitem{duplantier} H. Saleur and B. Duplantier,
Phys. Rev. Lett. {\bf 58}, 2325 (1986).
                                             
\bibitem{Gardiner}C. W. Gardiner and S. Chatuverdi, J. Stat. Phys.
{\bf 17}, 429 (1977); ibid {\bf 18}, 501 (1978). See also
C. W. Gardiner, {\it Handbook of Stochastic Methods}, Springer-Verlag,
Berlin and Heidelberg, 1985.

\bibitem{Yo} M. A. Mu\~noz, Phys. Rev. E. {\bf 57}, 1377  (1998).

\bibitem{Peliti}  L. Peliti, J. Physique  {\bf 46}, 1469 (1985). 
See also, B. P. Lee and J. Cardy, J. Stat. Phys. {\bf 80}, 971 (1995).


\bibitem{mass} Observe that the critical value of the linear 
term coefficient is equal to zero only at mean field (tree) level. 
As soon as fluctuations, i.e. diagrammatic corrections, 
are taken into account the critical mass is shifted to 
a non-vanishing value \cite{epid,Jan}.

\bibitem{BTW} P. Bak, C. Tang and K. Wiesenfeld,
Phys. Rev. Lett. {\bf 59}, 381 (1987);
Phys. Rev. A {\bf 38}, 364 (1988).      

\bibitem{FES} 
 A. Vespignani,  R. Dickman, M. A.  Mu{\~n}oz, and S. Zapperi,
Phys. Rev. E {\bf 62}, 4564 (2000).           
       

\end{references}
\end{document}